%% file: main.tex
\documentclass[sigconf, screen, nonacm]{acmart}
\AtBeginDocument{%
  }

\setcopyright{acmlicensed}
\copyrightyear{2025}
\acmYear{2025}
\acmDOI{XXXXXXX.XXXXXXX}
\acmConference[UIST '25]{ACM Symposium on User Interface Software and Technology}{September 28--October 1, 2025}{Busan, Korea}
\acmBooktitle{UIST '25: ACM ACM Symposium on User Interface Software and Technology, September 28--October 1, 2025, Busan, Korea} 
\acmISBN{978-1-4503-XXXX-X/25/09}

\usepackage{balance}
\usepackage{xcolor}
\usepackage{pgffor}
\usepackage[most]{tcolorbox} 
\usepackage{graphicx}       
\usepackage{booktabs}       
\usepackage{amsmath}        
\usepackage{hyperref}       
\usepackage{tikz}
\usepackage{forest}
\usepackage{tabularx}

\newtcolorbox{designRecom}[2][]{%
    enhanced,
    breakable,
    colback=blue!3,          
    colframe=blue!40!black,  
    coltitle=white,          
    colbacktitle=blue!30!black, 
    boxrule=0.6pt,           
    arc=3pt,                 
    fontupper=\small, 
    left=8pt, right=8pt,      
    top=6pt, bottom=6pt,
    title=\textbf{Design Recommendation:}\\\textnormal{#2},
    #1
}

\begin{document}

\title[Design Space of AI-assisted Research Tools]{The Design Space of Recent AI-assisted Research Tools for Ideation, Sensemaking, and Scientific Creativity}

\author{Runlong Ye}
\orcid{0000-0003-1064-2333}
\email{harryye@cs.toronto.ca}
\affiliation{
  \institution{Computer Science, \\University of Toronto}
  \city{Toronto}
  \state{Ontario}
  \country{Canada}
}

\author{Matthew Varona}
\orcid{0009-0005-6201-973X}
\email{varona@cs.toronto.edu}
\affiliation{
  \institution{Computer Science, \\University of Toronto}
  \city{Toronto}
  \state{Ontario}
  \country{Canada}
}

\author{Oliver Huang}
\orcid{0009-0007-1585-1229}
\email{oliver@cs.toronto.edu}
\affiliation{
  \institution{Computer Science, \\University of Toronto}
  \city{Toronto}
  \state{Ontario}
  \country{Canada}
}

\author{Patrick Yung Kang Lee}
\orcid{0000-0002-3385-5756}
\email{patricklee@cs.toronto.edu}
\affiliation{
  \institution{Computer Science, \\University of Toronto}
  \city{Toronto}
  \state{Ontario}
  \country{Canada}
}

\author{Michael Liut}
\orcid{0000-0003-2965-5302}
\email{michael.liut@utoronto.ca}
\affiliation{
  \institution{Mathematical and Computational Sciences, \\University of Toronto Mississauga}
  \city{Mississauga}
  \state{Ontario}
  \country{Canada}
}

\author{Carolina Nobre}
\orcid{0000-0002-2892-0509}
\email{cnobre@cs.toronto.edu}
\affiliation{
  \institution{Computer Science, \\University of Toronto}
  \city{Toronto}
  \state{Ontario}
  \country{Canada}
}


\begin{abstract}
\input{0-abstract}
\end{abstract}



\maketitle

\input{1-introduction}
\input{2-related-work}
\input{3-method}
\input{4-design-space}
\input{5-result}
\input{6-discussion}


\bibliographystyle{ACM-Reference-Format}
\balance
\bibliography{reference}

\appendix
\input{99-appendix}

\end{document}

%% file: 0-abstract.tex
Generative AI (GenAI) tools are radically expanding the scope and capability of automation in knowledge work such as academic research. While promising for augmenting cognition and streamlining processes, AI-assisted research tools may also increase automation bias and hinder critical thinking. To examine recent developments, we surveyed publications from leading HCI venues over the past three years, closely analyzing thirteen tools to better understand the novel capabilities of these AI-assisted systems and the design spaces they enable: seven employing traditional AI or customized transformer-based approaches, and six integrating open-access large language models (LLMs). Our analysis characterizes the emerging design space, distinguishes between tools focused on workflow mimicry versus generative exploration, and yields four critical design recommendations to guide the development of future systems that foster meaningful cognitive engagement: providing user agency and control, differentiating divergent/convergent thinking support, ensuring adaptability, and prioritizing transparency/accuracy. This work discusses how these insights signal a shift from mere workflow replication towards generative co-creation, presenting new opportunities for the community to craft intuitive, AI-driven research interfaces and interactions.     


%% file: 1-introduction.tex
\section{Introduction}
Generative AI (GenAI) is transforming the landscape of research and creative workflows by radically expanding the scope and capabilities of automation. Recent advancements in large language models (LLMs) are redefining how researchers engage in ideation, sensemaking, and scientific creativity. At the same time, integrating GenAI into research workflows raises critical questions about its effects on human cognition. Although these systems promise to augment human intelligence and streamline research processes, evidence indicates overreliance on automated output can reduce critical thinking and increase automation bias \cite{lee2025impact,passi2022overreliance}. Furthermore, concerns have been raised about the potential of GenAI to steer users too heavily, thus diminishing human capacity for interpretation and reflection \cite{10.1145/3544548.3581066}. 

However, the landscape of AI research tools is broader than LLM-powered systems alone. Other recent tools utilize customized transformers, such as fine-tuned BERT models or rule-based transformer variants, to provide deterministic, workflow-aligned assistance. These transformer-based models differ fundamentally from general-purpose LLMs like \texttt{GPT-4} by offering constrained and transparent AI assistance aligned closely with researchers' existing cognitive processes.

The latest developments in the AI industry further emphasize these trends: OpenAI's Deep Research \cite{openai2025deepresearch}, an AI agent capable of synthesizing information autonomously to generate comprehensive research reports, highlights the rapid push toward knowledge work automation. Soon after, Google's AI co-scientist \cite{google2025aicoScientist}, a multi-agent system for iterative hypothesis refinement, illustrates the expanding role of AI in scientific discovery.

Addressing the challenges posed by these trends, a human-centered approach to AI~\cite{shneiderman2020human} emerges as a promising strategy. This approach leverages the computational power of AI while explicitly preserving and fostering human roles in guiding, interpreting, and refining AI output, thus sustaining deeper cognitive engagement.

To address this gap, we investigate recent AI-assisted research tools specifically designed to support ideation, sensemaking, and scientific creativity. While existing reviews, such as Pang et al.'s work on LLM usage in CHI \cite{pang2025understandingllmificationchiunpacking}, cover broad trends, our analysis concerns the specific design space of research-oriented AI tools, including specialized systems beyond general-purpose LLMs (like those using customized transformers or traditional AI). We examine their distinct influence on research workflows and cognitive engagement, aiming to uncover critical design considerations for fostering effective human-AI partnerships in research.

Through our analysis, we identify four key dimensions that influence cognitive engagement with AI-powered research tools: user agency and control (\autoref{sec:agency}); divergent and convergent thinking (\autoref{sec:divergent-convergent}); adaptability (\autoref{sec:adaptability}); and accuracy (\autoref{sec:accuracy}). By situating existing tools within this multidimensional framework, we characterize critical design choices, highlight potential cognitive pitfalls, and reveal important distinctions between GenAI systems, which typically guide users toward predefined pathways of discovery, and more traditional or customized transformer-based tools that closely mimic existing workflows and emphasize user autonomy (\autoref{sec:genai-vs-ai}).

Articulating these distinctions opens opportunities to design advanced systems for research co-creation (\autoref{sec:oppotunities}), aimed at empowering researchers. Thus, our work clarifies the emerging design space by contrasting different AI approaches, offering concrete insights for designing future tools that prioritize researcher agency and encourage meaningful cognitive engagement in the research workflow.

%% file: 2-related-work.tex
\section{Related Work}
Our analysis of AI-assisted research tools is informed by prior research in augmenting human intellect, computational support for cognitive tasks like sensemaking and creativity, evolving paradigms of human-AI interaction, and recent systematic reviews mapping AI's influence in research.
\subsection{AI Augmentation in Research and Knowledge Work}
The vision of computationally augmenting human intellect, famously articulated by \citet{bush1945we} and demonstrated by pioneers like Engelbart \cite{engelbart2023augmenting}, laid the groundwork for using technology to enhance knowledge work. Early HCI efforts focused on structured information, interaction modalities, and information visualization as means to amplify cognition \cite{card1999readings}. While traditional AI offers expert systems and information retrieval, recent advances in AI, especially large-scale models, enable qualitatively different interactions, supporting open-ended generation, synthesis, and dialogue directly within research workflows \cite{birhane2023science}. This shift opens new possibilities for tools that, beyond retrieving information, actively participate in its interpretation and creation, demanding new interaction and system design approaches. Our work focuses specifically on the design space emerging from these recent, more capable AI assistants for research tasks.

\subsection{Computational Tools for Sensemaking and Creativity}
Sensemaking involves iterative processes of information foraging, structuring, and developing insights \cite{pirolli2005sensemaking}. HCI researchers have explored various computational supports, from visual analytics dashboards aiding exploration \cite{heer2005prefuse} to note-taking systems facilitating organization. AI introduces automation possibilities like theme clustering \cite{scholastic, sensemate} and summarization \cite{synergi}, but raises questions about how to design interfaces that effectively integrate these capabilities while supporting the user's cognitive process \cite{russell1993cost}. Addressing this challenge, recent systems like Sensecape explore novel interfaces that move beyond linear interaction; Sensecape, for instance, employs multilevel abstraction and enables fluid switching between foraging and sensemaking to support complex, non-linear sensemaking workflows within an LLM-powered environment \cite{Sensecape}. 

Similarly, creativity support tools (CSTs) aim to enhance human innovation \cite{shneiderman2007creativity}, often targeting divergent and convergent thinking.  While prior tools used techniques like computational analogy or constraint manipulation, modern AI enables approaches such as analogical retrieval at scale using advanced algorithms \cite{AnalogicalSearchEngine}. and GenAI specifically introduces novel mechanisms like prompt-based exploration \cite{IdeaSynth, disciplink}, creating a distinct design space for AI-powered CSTs that we analyze.

\subsection{Human-AI Collaboration Paradigms}
Effective integration of AI requires moving beyond automation towards genuine human-AI collaboration, often conceptualized as mixed-initiative interaction where both human and AI can contribute proactively \cite{10.1145/302979.303030}. Designing such systems necessitates balancing user agency and control with AI proactivity \cite{parasuraman2000model}. Over-reliance and automation bias \cite{lee2025impact, passi2022overreliance, 10.1145/3544548.3581066},  are significant risks. For instance, recent large-scale experiments found that while LLM assistance boosted performance during creative tasks, it could subsequently hinder participants' independent divergent and convergent thinking abilities when the AI was removed \cite{kumar2024human}. These findings stress the need for transparency, explainability, and user control mechanisms, core tenets of human-centered AI (HCAI) \cite{shneiderman2020human}. Building trust is crucial, requiring not just accurate AI outputs but also understandable processes \cite{lee2004trust}. The demand for transparency links directly to explainable AI (XAI) research \cite{arrieta2020explainable}, which seeks methods to make AI decisions interpretable. These foundational human-AI interaction challenges regarding agency, control, accuracy, and transparency directly inform the design dimensions explored in this paper's analysis.

\subsection{Systematic Analyses of LLMs in HCI and Research}
Systematic reviews are increasingly mapping the integration of Artificial Intelligence into HCI research. Notably,  \citet{pang2025understandingllmificationchiunpacking} systematically analyzed how Large Language Models (LLMs) have been utilized within recent CHI publications (2020–2024) Their review categorizes the application domains, identifies common roles played by LLMs (e.g., as system engines, research tools, or simulated users), and highlights methodological and ethical concerns. While this analysis provides a valuable overview of LLM adoption in HCI, our work differs significantly in two key aspects. First, we examine AI-powered research tools more broadly (including non-LLM techniques). And second, our analysis uniquely emphasizes cases where the AI-assisted tool itself is the primary research contribution, rather than studies that simply employ AI as an auxiliary methodological tool. This targeted perspective enables a deeper understanding of how these specialized systems are designed, evaluated, and contribute distinctly to research processes such as ideation, sensemaking, and scientific creativity. Our focused analysis complements broader reviews, like  \citet{luo2025llm4srsurveylargelanguage}, which survey the general use of AI/LLMs across all stages of the scientific research cycle.

%% file: 3-method.tex
\section{Method}

\input{table/papers}

This paper analyzes a selection of recent AI-assisted \emph{creative research tool}. Our scope specifically targets systems designed to aid the \textbf{co-creation of concepts and ideas} during the research process, distinguishing them from AI-assisted \textit{writing tools} which primarily focus on improving stylistic or rhetorical choices in written text. Based on this criterion, we identified thirteen representative systems papers published in top-tier HCI venues (i.e., CHI, CSCW, UIST, and ToCHI) over the last three years (2022-2024)\footnote{We include a relevant CHI 2025 paper made available on arXiv.}; details can be found in Table~\ref{tab:tool_classification}. 

This time period was selected as it reflects significant growth in the GenAI/LLM adoption, post-ChatGPT \cite{openai2022chatgpt}, capturing relevant and impactful developments in GenAI-driven research tools. Of the thirteen systems' papers surveyed, four of the systems represent a traditional AI approach, employing non-LLM machine learning techniques (e.g., Seq2Seq, RNN, topic modeling). Three of the recent systems utilize customized transformer-based models, providing task-specific, fine-tuned, or rule-based transformer architectures to aid researchers deterministically and transparently in research tasks. Additionally, six systems integrate open-access LLM-based functionalities (primarily using GPT-3 or similar models).



Categorizing the surveyed tools into traditional AI, customized transformer-based, and open-access LLM approaches allows us to effectively contrast their respective capabilities and design considerations. This comparative structure clarifies the fundamental shifts introduced by recent generative AI tools relative to more established AI paradigms.

The thematic dimensions presented in Section~\ref{sec:design-space} resulted from extensive discussions between the authors and consultation with an external expert specializing in AI-supported creativity and sensemaking. This collaborative and iterative approach produced the four critical dimensions addressed in the next section.

\subsection{Limitation}
While we aimed to capture influential recent developments, this review does not constitute a formal systematic literature review (SLR). Our selection process, focused on specific HCI venues and recent years, sought to identify a representative variety of AI-assisted research tools for ideation, sensemaking, and creativity across different underlying technologies (traditional AI, customized transformers, LLMs). However, this approach means our corpus may not be exhaustive, and other relevant tools published in different venues or using alternative terminology might have been missed. The findings should therefore be interpreted as reflecting trends and design dimensions observed within this specific, curated set of recent systems.

%% file: table/papers.tex
\begin{table*}[ht]
  \centering
  \small
  \begin{tabularx}{\textwidth}{l c X X l l}
    \toprule
    \textbf{Tool} & \textbf{Year} & \textbf{AI Usage} & \textbf{Research Support Area} & \textbf{Category} & \textbf{Reference} \\
    \midrule
    Analogical Search Engine & 2022 & Token-level ranking algorithm & Literature search \& creativity & Non-LLM & \cite{AnalogicalSearchEngine} \\
    Relatedly               & 2023 & Heuristic semantic matching & Literature review scaffolding & Non-LLM & \cite{Relatedly} \\
    Scholastic              & 2022 & Graph clustering \& visualization & Qualitative data analysis & Non-LLM & \cite{scholastic} \\
    Threddy                 & 2022 & NLP pipeline (parsing \& LDA) & Literature organization & Non-LLM & \cite{threddy} \\
    \midrule
    CoAICoder               & 2023 & Fine-tuned BERT for code suggestion & Collaborative qualitative analysis & Custom Transformer-based & \cite{CoAIcoder} \\
    PaTAT                   & 2023 & Fine-tuned BERT for interactive rule synthesis & Qualitative data analysis & Custom Transformer-based & \cite{PaTAT} \\
    SenseMate               & 2024 & Fine-tuned BERT for rationale extraction & Qualitative data analysis & Custom Transformer-based & \cite{sensemate} \\
    \midrule
    CollabCoder             & 2024 & GPT-family LLM & Collaborative qualitative analysis & Open-Access LLM-based & \cite{CollabCoder} \\
    CoQuest                 & 2024 & GPT-family LLM & Research question generation & Open-Access LLM-based & \cite{CoQuest} \\
    DiscipLink              & 2024 & GPT-family LLM & Interdisciplinary exploration & Open-Access LLM-based & \cite{disciplink} \\
    IdeaSynth               & 2025 & GPT-family LLM & Research idea generation & Open-Access LLM-based & \cite{IdeaSynth} \\
    PaperWeaver             & 2024 & GPT-family LLM & Literature recommendation & Open-Access LLM-based & \cite{paperweaver} \\
    Synergi                 & 2023 & GPT-family LLM and citation graph analytics & Literature synthesis \& sensemaking & Open-Access LLM-based & \cite{synergi} \\
    \bottomrule
  \end{tabularx}
  \caption{Classification of Recent AI-assisted Research Tools}
  \Description{
  The table provides a classification of the thirteen recent AI-assisted research tools analyzed in this paper. For each tool, the table lists its name, publication year, the specific AI technique it uses, the research area it supports, a category indicating the type of AI (Non-LLM, Custom Transformer-based, or Open-Access LLM-based), and the corresponding reference number. The table presents four Non-LLM tools first, followed by three Custom Transformer-based tools, and finally six Open-Access LLM-based tools. This organization allows comparison across different types of AI systems regarding their usage, support area, and publication timeline from 2022 to 2024 (with the exception of a CHI 2025 paper).
  }
  \label{tab:tool_classification}
\end{table*}

%% file: 4-design-space.tex
\section{Design Space}
\label{sec:design-space}


A fundamental challenge in integrating AI-powered research tools is ensuring that users remain cognitively ``in the loop'' rather than passively accepting AI-generated output. While automation promises efficiency and rapid idea generation, it also risks encouraging users to rely too heavily on system outputs, which can diminish human critical thinking and decision-making processes. We explore four design dimensions critical to navigating this tension.

\subsection{User Agency and Control}
\label{sec:agency}

Human-AI collaboration systems aim to improve productivity and creativity by offloading certain tasks from humans to AI systems while keeping the user in the driver's seat. Providing user agency and control via source material participation, the ability to refine AI output, and the ability to reject and override automated actions are crucial to maintaining cognitive engagement from the user.

\subsubsection{Engagement with Source Material.} For tools aimed at making sense of existing text content (such as qualitative coding, thematic analysis, or literature reviews), there is a delicate balance to be struck between AI assistance and user agency. Although recent advancements in LLM capabilities show promise in processing and summarizing large amounts of text, users still have to engage with the source text to build their own understanding and avoid model overreliance. One way to foster engagement is for user-selected text highlights to drive content generation. Systems such as Synergi \cite{synergi} and Threddy \cite{threddy} integrate PDF readers that transform user-selected highlights into seeds for AI-driven research thread generation, while Relatedly~\cite{Relatedly} further refine this process by organizing and highlighting overlapping research themes. Meanwhile, platforms like Scholastic \cite{scholastic},  SenseMate \cite{sensemate}, and PaTAT \cite{PaTAT} scaffold existing analysis methods with cluster suggestions, strategic sampling, or explainable rule-based coding aids, enabling AI assistance within user-controlled actions.

\subsubsection{Refining AI Output.} Human-centric AI tool design assumes the user is the expert, giving them the power to modify the output provided by the AI system. Most tools we reviewed allow users to edit and refine AI-generated artifacts to ensure that they meet the user's goals. In Threddy \cite{threddy}, for example, users can manually clean up errors in references and links extracted from a paper snippet. Beyond fixing errors, user editing can be designed into human-AI sensemaking systems to varying degrees. Arranging AI outputs into a more interpretable structure (such as node-link diagrams \cite{IdeaSynth, CoQuest} or outlines \cite{synergi}) can foster deeper engagement with suggestions. Finally, iterating on prompts and queries can help users gradually incorporate new ideas and discoveries into AI output. This is notably useful in cases such as the Analogical Search Engine \cite{AnalogicalSearchEngine}, where users may not know exactly how to initially prompt but can gain and apply new information with each re-prompt.

\subsubsection{Rejection and Overriding.} One aspect of control that merits further consideration is the ability of users to reject, override, or ignore model output. Rejection can manifest in systems implicitly. For example, editing or curating AI output implies rejection of the original content in part or whole. However, most of the research papers we reviewed had AI assistance embedded in the system, with minimal ability to ``turn off'' AI suggestions. One notable exception is SenseMate, which explicitly aims to provide AI theme suggestions on demand rather than by default\cite{sensemate}. In SenseMate, AI suggestions are hidden by default; users can also see the reasoning for the theme suggestions and explicitly reject them. Another example is PaTAT \cite{PaTAT}, which generates interpretable rule-based code suggestions that researchers can inspect and choose to accept or reject, ensuring the AI remains an optional aide. Likewise, Scholastic's text clustering algorithm does not impose keywords on clusters. Instead, users have the option of either developing internal mental models of the meaning of each cluster or providing the algorithm with explicit codes that can be iterated on \cite{scholastic}. In all these cases, the AI's role is advisory rather than authoritative, allowing researchers to override or ignore recommendations that do not fit their needs.

\begin{designRecom}{Provide User Agency and Control}
\textit{Design Insight:} Users should have meaningful control and agency within AI-assisted systems to ensure that they remain the primary decision-makers. Users should be able to engage with the source text and edit, refine, or reject AI output.
\smallskip

\textit{Implications:} Future systems should consider
\begin{itemize} 
\item \textbf{Active Engagement with Source Material:} AI tools must encourage users to interact with underlying data (e.g., through highlights or annotations) so that understanding is built from the ground up.
\item \textbf{Editable and Iterative Output:} Systems should allow and even invite users to refine AI outputs, shaping them to their context.
\item \textbf{Override Mechanisms:} Provide easy ways for users to bypass or disable AI suggestions, preserving human judgment as the final arbiter.
\end{itemize}

\end{designRecom}

\subsection{Divergent and Convergent Thinking}
\label{sec:divergent-convergent}

\subsubsection{Divergent Thinking}
 Divergent thinking is essential for expanding the horizons of research. AI tools support this by generating exploratory questions and novel insights that researchers may not have thought about. For example, the Analogical Search Engine~\cite{AnalogicalSearchEngine}, DiscipLink~\cite{disciplink}, and IdeaSynth~\cite{IdeaSynth} provide functionalities to explore creative connections in seemingly unrelated fields. However, when using an AI tool for divergent thinking in isolation, this approach can overwhelm users with many novel ideas without providing the guidance necessary to narrow them down or refine them effectively. This highlights the importance of coupling divergent thinking features with mechanisms that support reflection, prioritization, or convergence. Without such scaffolding, the creative potential of these systems may be underutilized or lead to cognitive overload.
 
\subsubsection{Convergent Thinking}
In convergent thinking, AI tools transform complex data into clear and structured insights, acting as intelligent partners that help researchers distill complex information into coherent narratives. For example, SenseMate \cite{sensemate} generates transparent, rationale-driven theme recommendations, helping users deeply engage with the source material. PaTAT similarly offers transparent, rule-based, coding suggestions, facilitating thematic analysis in a controlled manner \cite{PaTAT}. Focusing on the same task, CollabCoder \cite{CollabCoder} utilizes LLMs to automatically generate qualitative code suggestions and facilitate structured group discussions, thereby bridging individual insights into a collective consensus. Complementing these approaches, Scholastic~\cite{scholastic} employs advanced visual analytics to help teams organize and interpret complex datasets, while PaperWeaver~\cite{paperweaver} presents contextualized links that highlight the most relevant insights. These convergent functionalities make thematic grouping and filtering more efficient and transparent. However, without opportunities for user reflection or iterative refinement, convergent tools may risk reinforcing surface-level interpretations. Therefore, effective design should balance automated structuring with user agency to ensure meaningful synthesis rather than passive consumption.

\subsubsection{Mixed-Thinking}
Some tools strike a balance by supporting both divergent and convergent thinking. Threddy~\cite{threddy}, for example, allows users to input various requests so that LLM systems can organize ideas into coherent themes. They also leverage hierarchical structures to discover new connections based on user input. Similarly, Synergi~\cite{synergi} uses citation graphs and language models to expand research threads and consolidate them. However, these mixed-thinking approaches often leave users with limited control over the balance between exploration and refinement, which may hinder effective sensemaking. While too much convergence can stifle creative exploration, excessive divergence may lead to cognitive overload. A well-designed research tool should flexibly support both modalities, but must also make these modes visible and adjustable to the user. Providing intuitive controls over the level and type of AI assistance empowers researchers to dynamically shift between exploratory and analytical workflows, preserving autonomy while enhancing insight generation.

\begin{designRecom}{Differentiating Divergent and Convergent Thinking}
\textit{Design Insight:} Users require distinct forms of AI support aligned with different phases of the research process. Generative, divergent capabilities are particularly valuable during early-stage ideation, while convergent tools that structure, filter, or synthesize information become essential during analysis and synthesis. Designing effective AI assistance requires differentiating support for these distinct divergent and convergent thinking modes.
\smallskip

\textit{Implications:} Future systems should consider:

\begin{itemize}
\item \textbf{Transparency of AI Contributions:} Clearly communicate whether the AI is supporting divergent exploration, convergent refinement, or a hybrid approach to help users calibrate their expectations and interactions. 
\item \textbf{User Control Over AI Assistance:} Allow users to modulate the level and type of AI involvement dynamically, adjusting the balance between creativity and structure based on task context and personal workflow preferences.
\item \textbf{Support for Mode Transitions:} Provide scaffolding or UI cues that help users shift between divergent and convergent thinking modes without cognitive friction or loss of continuity.  
\end{itemize}

\end{designRecom}

\subsection{Adaptability}
\label{sec:adaptability}

Adaptability in AI-assisted research tools refers to a system's capacity to support the diversity of tasks, workflows, and preferences of researchers. Ways this could be supported in tooling include flexible input mechanisms, fluid and non-linear workflows, and context-sensitive design.

\subsubsection{Flexible Input \& Customization} Systems that prioritize flexible input mechanisms empower users to tailor the tool's behavior from the outset. For example, the Analogical Search Engine~\cite{AnalogicalSearchEngine} leverages a custom ranking algorithm that focuses on user-defined ``purposes'' and ``mechanisms'' to modulate search results according to varying research objectives, allowing researchers to specify the kind of analogical relationships they seek. Similarly, DiscipLink \cite{disciplink} offers flexibility by letting users explicitly adjust interdisciplinary search parameters, guiding the AI to explore cross-domain connections relevant to specific research interests. Such customizable interactions highlight the importance of flexibility, enabling users to align AI behavior closely with their investigative goals, thereby preserving user agency and enhancing the relevance of system outputs.

\subsubsection{Fluid, Non-Linear Workflows} A second adaptability factor is found in tools that support non-linear, iterative workflows. The RQ Flow Editor in CoQuest~\cite{CoQuest} symbolizes this by eschewing predefined categories and instead promoting continuous refinement of ideas through a mixed-initiative interaction where an AI agent suggests new research questions and users can provide feedback in a loop. Likewise, IdeaSynth~\cite{IdeaSynth} supports both literature-driven and idea-driven explorations through dynamic facet generation and prompt customization, enabling users to decompose an initial idea into finer-grained aspects and explore variations of them. Tools like Scholastic~\cite{scholastic} further demonstrate adaptability by allowing researchers to shift seamlessly between exploration, strategic sampling, and coding via an interactive document viewer and word clustering interface, which highlights the value of fluid transitions in non-linear research processes.

\subsubsection{Mixed-Initiative \& Context-Sensitive Design} 
Adaptability also manifests in systems that accommodate varied research approaches through mixed-initiative interactions and context-sensitive features. For instance, Synergi~\cite{synergi} offers a mixed-initiative workflow that caters to both bottom-up and top-down strategies, allowing users to engage with content according to their preferred mode of inquiry and seamlessly combine machine-generated summaries and user-curated threads. Interdisciplinary platforms like DiscipLink~\cite{disciplink} contribute to this design space by facilitating the generation and refinement of research questions across multiple disciplines, albeit with some limitations regarding persistent user customization.

\begin{designRecom}{Ensure Adaptability and Workflow Flexibility} \textit{Design Insight:}
Research tools must provide flexible workflows and customizable interfaces to address varied researcher needs and expertise levels. An adaptive system can accommodate both novices (by guiding them step-by-step or providing explainable suggestions) and experts (by allowing rapid, free-form interaction), as well as different research methodologies.

\smallskip

\textit{Implications:} Future systems should consider:
\begin{itemize}
\item \textbf{Flexible Input \& Customization:}
Enable users to tailor key parameters and input formats to align system behavior with their individual research goals.
\item \textbf{Fluid, Non-Linear Workflows:} Provide multiple entry points and modular interfaces that allow researchers to navigate, reorganize, and iteratively refine their inquiry as new insights emerge.
\item \textbf{Mixed-Initiative \& Context-Sensitive Design:} Implement context-aware features that respond to shifts in research focus, ensuring a balanced mix of automated support and user control.
\end{itemize}
\end{designRecom}

\subsection{Accuracy} 
\label{sec:accuracy} 
Ensuring that users receive accurate, unbiased, and contextually relevant information is paramount in AI-assisted research tools. This is especially challenging as systems integrate powerful generative models: issues such as hallucination and contextual drift inherent to large language models can undermine trust and usefulness. In response, researchers have developed various strategies that combine human oversight, contextual grounding, and carefully managed efficiency-accuracy trade-offs.

\subsubsection{Interactive Interfaces for Accuracy Validation} A key design pattern embeds interactive mechanisms that enable real-time verification of AI outputs by linking inferences directly to their original sources. For instance, SenseMate \cite{sensemate} employs a ``View Reason'' feature that highlights source phrases underlying a theme recommendation, thus promoting local explainability and inviting critical evaluation rather than passive acceptance. Similarly, Synergi~\cite{synergi} and PaperWeaver~\cite{paperweaver} enhance their LLM-generated summaries by providing citation contexts and contextualized descriptions. Synergi groups related information to offer clear reference, while PaperWeaver uses aspect-based summaries (e.g., problem, method, findings) alongside paper comparisons to help researchers quickly assess the relevance of new publications. In the domain of qualitative analysis, PaTAT [5] also prioritizes transparency: as users code data, PaTAT produces human-readable rules explaining its coding suggestions, enabling researchers to inspect how the AI-derived a suggestion before deciding to use it. By grounding AI outputs in traceable rationale or source material, these systems encourage users to verify information and reduce blind trust.

\subsubsection{Iterative Refinement and Human-in-the-Loop Strategies} Complementing interactive validation, iterative refinement processes further emphasize human oversight. Scholastic~\cite{scholastic} demonstrates this approach by utilizing a machine-in-the-loop framework for qualitative text coding, wherein user feedback continuously refines coding decisions through rationale extraction models that enhance transparency and trust. Likewise, Threddy~\cite{threddy} and IdeaSynth~\cite{IdeaSynth} empower researchers to actively shape AI outputs. Threddy facilitates the extraction and iterative organization of research threads, while IdeaSynth leverages LLMs to propose new ideas that users can further refine. Together, these strategies enable researchers to combine automated suggestions with their own expert judgment.

\begin{designRecom}{Prioritize Transparency, Accuracy, And Ways to Validate} \textit{Design Insight:} In scientific research, accuracy is non-negotiable. AI systems must balance automation with robust mechanisms for users to verify and understand AI outputs. Clear traceability from output to source or rationale is essential for trust, especially when using generative models that might introduce errors.

\textit{Implications:} Future systems should consider
\begin{itemize} 
\item \textbf{Transparent Rationale Explanations:} Allow users to understand the underlying reasoning behind AI outputs. 
\item \textbf{Citation Contexts:} Provide direct access to source materials, enabling quick cross-referencing and fact-checking of AI-generated content. 
\item \textbf{Human-in-the-Loop Reviews:} Facilitate iterative user feedback that mitigates risks such as hallucinations and inadvertent biases, ensuring that final outputs have been vetted or corrected by human experts.
\end{itemize}

\end{designRecom}

%% file: 5-result.tex
\section{A Tale of Two AIs: Workflow Mimicry vs. Generative Exploration}
\label{sec:genai-vs-ai}
As we surveyed recent AI-assisted research tools, we observed two distinct approaches: systems based on deterministic fine-tuned transformers (e.g., CoAICoder, PaTAT, SenseMate) and traditional non-LLM models (e.g., Analogical Search Engine, Relatedly, Threddy), and those leveraging open-access LLMs (e.g., CoQuest, IdeaSynth, Synergi). Both approaches enhance research but differ fundamentally: one emphasizes deterministic workflow mimicry, while the other encourages open-ended generative exploration.

Customized transformer-based and non-LLM research tools augment established research workflows by automating clearly defined tasks. The Analogical Search Engine retrieves literature using token-level similarity algorithms; PaTAT and SenseMate provide structured and interpretable coding suggestions directly embedded in qualitative workflows; CoAICoder supports team consensus via tailored coding suggestions. These tools reliably replicate manual processes, providing consistency and control.

In contrast, research tools that use open-access GPT-family language models focus on generative interactions, dynamically proposing novel insights or research directions beyond explicit user input. CoQuest spontaneously generates exploratory research questions; IdeaSynth injects diverse literature-grounded ideas; Synergi synthesizes insights through generative summarization. These tools introduce creative, non-deterministic contributions, requiring researchers to actively interpret and shape their AI-generated outputs.

This emerging division marks an important shift: customized transformers and non-LLM tools provide deterministic, structured support aligned with current methods, ensuring consistency and reliability. Conversely, LLM-enabled tools foster creative exploration, embracing uncertainty to catalyze discovery. The ongoing design challenge lies in balancing these philosophies: integrating the reliability and interpretability of deterministic AI assistance with the creativity and serendipity offered by GenAI. Future ``tools for thought'' might optimally combine the strengths of both paradigms, allowing researchers to harness AI not just for task efficiency but also for rich cognitive engagement, without sacrificing intellectual autonomy.

%% file: 6-discussion.tex



\section{Opportunities: Designing for Generative Co-Creation}
\label{sec:oppotunities}

The analysis highlights a clear developing trend in AI-assisted research tools: a move from automating established workflows towards enabling genuine human-AI co-creation, particularly with the rise of generative AI. This presents exciting opportunities for the UIST community to design the next generation of interfaces and interactions that harness GenAI's potential while fostering deep cognitive engagement:

\begin{enumerate}
    \item \textbf{Interactive Steering Between Exploration and Refinement:} Our findings highlight a tension between deterministic, workflow-mimicking tools and open-ended generative exploration. A key opportunity lies in designing novel interaction techniques that allow researchers to fluidly navigate between these modes. Imagine interfaces that enable users to dynamically adjust the AI's level of initiative (from structured assistance to proactive generation) or seamlessly transition AI outputs between exploratory canvases (supporting divergence) and structured analytical frameworks (supporting convergence), potentially using direct manipulation or adaptive UIs.
    \item \textbf{Tangible Mechanisms for Agency and Transparency:} The design dimensions identified (Agency, Thinking Modes, Adaptability, Accuracy) require concrete implementations in interfaces. Beyond simple editing or source-linking, future systems could incorporate interactive visualizations of AI uncertainty or provenance, explainable AI techniques tailored for generative models within the research context, and controls that allow users to actively guide the AI's reasoning process, perhaps by specifying constraints, providing critiques, or adjusting model parameters through intuitive interfaces. This moves transparency from a passive feature to an active, user-driven process.
    \item \textbf{Scaffolding Critical Engagement with Generative Outputs:} Addressing automation bias requires interfaces that don't just permit critical assessment but actively encourage it. Opportunities exist to design ``cognitive scaffolds'' within the UI, features that prompt users to question AI suggestions, compare alternatives generated under different assumptions, or articulate justifications for incorporating AI contributions. This involves designing for reflection and critical dialogue, not just task completion.
    \item \textbf{New Paradigms for Research Interaction:} GenAI unlocks fundamentally new ways for researchers to interact with information and ideas. Future work can explore beyond dialogue-based interfaces for co-developing hypotheses, multi-modal systems where users can fluidly move between text generation, data visualization, and code execution driven by AI, and agent-based systems where researchers can delegate exploratory tasks to AI collaborators who report back findings through interactive summaries.
\end{enumerate}

By focusing on these interface and interaction challenges, the community can shape GenAI-powered tools that truly augment scientific creativity and insight, ensuring researchers remain firmly in control of the discovery process while benefiting from the power of generative AI partners.

%% file: 99-appendix.tex

%% file: main.bbl

\begin{thebibliography}{36}


\ifx \showCODEN    \undefined \def \showCODEN     #1{\unskip}     \fi
\ifx \showISBNx    \undefined \def \showISBNx     #1{\unskip}     \fi
\ifx \showISBNxiii \undefined \def \showISBNxiii  #1{\unskip}     \fi
\ifx \showISSN     \undefined \def \showISSN      #1{\unskip}     \fi
\ifx \showLCCN     \undefined \def \showLCCN      #1{\unskip}     \fi
\ifx \shownote     \undefined \def \shownote      #1{#1}          \fi
\ifx \showarticletitle \undefined \def \showarticletitle #1{#1}   \fi
\ifx \showURL      \undefined \def \showURL       {\relax}        \fi
\providecommand\bibfield[2]{#2}
\providecommand\bibinfo[2]{#2}
\providecommand\natexlab[1]{#1}
\providecommand\showeprint[2][]{arXiv:#2}

\bibitem[ope(2025)]%
        {openai2025deepresearch}
 \bibinfo{year}{2025}\natexlab{}.
\newblock \bibinfo{title}{Introducing Deep Research}.
\newblock \bibinfo{howpublished}{\url{https://openai.com/index/introducing-deep-research}}.
\newblock
\newblock
\shownote{Accessed: 2025-02-20}.


\bibitem[Arrieta et~al\mbox{.}(2020)]%
        {arrieta2020explainable}
\bibfield{author}{\bibinfo{person}{Alejandro~Barredo Arrieta}, \bibinfo{person}{Natalia D{\'\i}az-Rodr{\'\i}guez}, \bibinfo{person}{Javier Del~Ser}, \bibinfo{person}{Adrien Bennetot}, \bibinfo{person}{Siham Tabik}, \bibinfo{person}{Alberto Barbado}, \bibinfo{person}{Salvador Garc{\'\i}a}, \bibinfo{person}{Sergio Gil-L{\'o}pez}, \bibinfo{person}{Daniel Molina}, \bibinfo{person}{Richard Benjamins}, {et~al\mbox{.}}} \bibinfo{year}{2020}\natexlab{}.
\newblock \showarticletitle{Explainable Artificial Intelligence (XAI): Concepts, taxonomies, opportunities and challenges toward responsible AI}.
\newblock \bibinfo{journal}{\emph{Information fusion}}  \bibinfo{volume}{58} (\bibinfo{year}{2020}), \bibinfo{pages}{82--115}.
\newblock


\bibitem[Birhane et~al\mbox{.}(2023)]%
        {birhane2023science}
\bibfield{author}{\bibinfo{person}{Abeba Birhane}, \bibinfo{person}{Atoosa Kasirzadeh}, \bibinfo{person}{David Leslie}, {and} \bibinfo{person}{Sandra Wachter}.} \bibinfo{year}{2023}\natexlab{}.
\newblock \showarticletitle{Science in the age of large language models}.
\newblock \bibinfo{journal}{\emph{Nature Reviews Physics}} \bibinfo{volume}{5}, \bibinfo{number}{5} (\bibinfo{year}{2023}), \bibinfo{pages}{277--280}.
\newblock


\bibitem[Bush et~al\mbox{.}(1945)]%
        {bush1945we}
\bibfield{author}{\bibinfo{person}{Vannevar Bush} {et~al\mbox{.}}} \bibinfo{year}{1945}\natexlab{}.
\newblock \showarticletitle{As we may think}.
\newblock \bibinfo{journal}{\emph{The atlantic monthly}} \bibinfo{volume}{176}, \bibinfo{number}{1} (\bibinfo{year}{1945}), \bibinfo{pages}{101--108}.
\newblock


\bibitem[Card et~al\mbox{.}(1999)]%
        {card1999readings}
\bibfield{author}{\bibinfo{person}{Stuart~K Card}, \bibinfo{person}{Jock Mackinlay}, {and} \bibinfo{person}{Ben Shneiderman}.} \bibinfo{year}{1999}\natexlab{}.
\newblock \bibinfo{booktitle}{\emph{Readings in information visualization: using vision to think}}.
\newblock \bibinfo{publisher}{Morgan Kaufmann}.
\newblock


\bibitem[Engelbart(1962)]%
        {engelbart2023augmenting}
\bibfield{author}{\bibinfo{person}{Douglas~C Engelbart}.} \bibinfo{year}{1962}\natexlab{}.
\newblock \showarticletitle{Augmenting human intellect: A conceptual framework}.
\newblock In \bibinfo{booktitle}{\emph{Augmented Education in the Global Age}}. \bibinfo{publisher}{Routledge}, \bibinfo{pages}{13--29}.
\newblock


\bibitem[Gao et~al\mbox{.}(2023)]%
        {CoAIcoder}
\bibfield{author}{\bibinfo{person}{Jie Gao}, \bibinfo{person}{Kenny Tsu~Wei Choo}, \bibinfo{person}{Junming Cao}, \bibinfo{person}{Roy Ka-Wei Lee}, {and} \bibinfo{person}{Simon Perrault}.} \bibinfo{year}{2023}\natexlab{}.
\newblock \showarticletitle{CoAIcoder: Examining the Effectiveness of AI-assisted Human-to-Human Collaboration in Qualitative Analysis}.
\newblock \bibinfo{journal}{\emph{ACM Trans. Comput.-Hum. Interact.}} \bibinfo{volume}{31}, \bibinfo{number}{1}, Article \bibinfo{articleno}{6} (\bibinfo{date}{Nov.} \bibinfo{year}{2023}), \bibinfo{numpages}{38}~pages.
\newblock
\showISSN{1073-0516}
\href{https://doi.org/10.1145/3617362}{doi:\nolinkurl{10.1145/3617362}}


\bibitem[Gao et~al\mbox{.}(2024)]%
        {CollabCoder}
\bibfield{author}{\bibinfo{person}{Jie Gao}, \bibinfo{person}{Yuchen Guo}, \bibinfo{person}{Gionnieve Lim}, \bibinfo{person}{Tianqin Zhang}, \bibinfo{person}{Zheng Zhang}, \bibinfo{person}{Toby Jia-Jun Li}, {and} \bibinfo{person}{Simon~Tangi Perrault}.} \bibinfo{year}{2024}\natexlab{}.
\newblock \showarticletitle{CollabCoder: A Lower-barrier, Rigorous Workflow for Inductive Collaborative Qualitative Analysis with Large Language Models}. In \bibinfo{booktitle}{\emph{Proceedings of the 2024 CHI Conference on Human Factors in Computing Systems}} (Honolulu, HI, USA) \emph{(\bibinfo{series}{CHI '24})}. \bibinfo{publisher}{Association for Computing Machinery}, \bibinfo{address}{New York, NY, USA}, Article \bibinfo{articleno}{11}, \bibinfo{numpages}{29}~pages.
\newblock
\showISBNx{9798400703300}
\href{https://doi.org/10.1145/3613904.3642002}{doi:\nolinkurl{10.1145/3613904.3642002}}


\bibitem[Gebreegziabher et~al\mbox{.}(2023)]%
        {PaTAT}
\bibfield{author}{\bibinfo{person}{Simret~Araya Gebreegziabher}, \bibinfo{person}{Zheng Zhang}, \bibinfo{person}{Xiaohang Tang}, \bibinfo{person}{Yihao Meng}, \bibinfo{person}{Elena~L. Glassman}, {and} \bibinfo{person}{Toby Jia-Jun Li}.} \bibinfo{year}{2023}\natexlab{}.
\newblock \showarticletitle{PaTAT: Human-AI Collaborative Qualitative Coding with Explainable Interactive Rule Synthesis}. In \bibinfo{booktitle}{\emph{Proceedings of the 2023 CHI Conference on Human Factors in Computing Systems}} (Hamburg, Germany) \emph{(\bibinfo{series}{CHI '23})}. \bibinfo{publisher}{Association for Computing Machinery}, \bibinfo{address}{New York, NY, USA}, Article \bibinfo{articleno}{362}, \bibinfo{numpages}{19}~pages.
\newblock
\showISBNx{9781450394215}
\href{https://doi.org/10.1145/3544548.3581352}{doi:\nolinkurl{10.1145/3544548.3581352}}


\bibitem[Gottweis et~al\mbox{.}(2025)]%
        {google2025aicoScientist}
\bibfield{author}{\bibinfo{person}{Juraj Gottweis}, \bibinfo{person}{Wei-Hung Weng}, \bibinfo{person}{Alexander Daryin}, \bibinfo{person}{Tao Tu}, \bibinfo{person}{Anil Palepu}, \bibinfo{person}{Petar Sirkovic}, \bibinfo{person}{Artiom Myaskovsky}, \bibinfo{person}{Felix Weissenberger}, \bibinfo{person}{Keran Rong}, \bibinfo{person}{Ryutaro Tanno}, \bibinfo{person}{Khaled Saab}, \bibinfo{person}{Dan Popovici}, \bibinfo{person}{Jacob Blum}, \bibinfo{person}{Fan Zhang}, \bibinfo{person}{Katherine Chou}, \bibinfo{person}{Avinatan Hassidim}, \bibinfo{person}{Burak Gokturk}, \bibinfo{person}{Amin Vahdat}, \bibinfo{person}{Pushmeet Kohli}, \bibinfo{person}{Yossi Matias}, \bibinfo{person}{Andrew Carroll}, \bibinfo{person}{Kavita Kulkarni}, \bibinfo{person}{Nenad Tomasev}, \bibinfo{person}{Yuan Guan}, \bibinfo{person}{Vikram Dhillon}, \bibinfo{person}{Eeshit~Dhaval Vaishnav}, \bibinfo{person}{Byron Lee}, \bibinfo{person}{Tiago R~D Costa}, \bibinfo{person}{José~R Penadés}, \bibinfo{person}{Gary Peltz},
  \bibinfo{person}{Yunhan Xu}, \bibinfo{person}{Annalisa Pawlosky}, \bibinfo{person}{Alan Karthikesalingam}, {and} \bibinfo{person}{Vivek Natarajan}.} \bibinfo{year}{2025}\natexlab{}.
\newblock \bibinfo{title}{Towards an AI co-scientist}.
\newblock
\showeprint[arxiv]{2502.18864}~[cs.AI]
\urldef\tempurl%
\url{https://arxiv.org/abs/2502.18864}
\showURL{%
\tempurl}


\bibitem[Heer et~al\mbox{.}(2005)]%
        {heer2005prefuse}
\bibfield{author}{\bibinfo{person}{Jeffrey Heer}, \bibinfo{person}{Stuart~K Card}, {and} \bibinfo{person}{James~A Landay}.} \bibinfo{year}{2005}\natexlab{}.
\newblock \showarticletitle{Prefuse: a toolkit for interactive information visualization}. In \bibinfo{booktitle}{\emph{Proceedings of the SIGCHI conference on Human factors in computing systems}}. \bibinfo{pages}{421--430}.
\newblock


\bibitem[Hong et~al\mbox{.}(2022)]%
        {scholastic}
\bibfield{author}{\bibinfo{person}{Matt-Heun Hong}, \bibinfo{person}{Lauren~A. Marsh}, \bibinfo{person}{Jessica~L. Feuston}, \bibinfo{person}{Janet Ruppert}, \bibinfo{person}{Jed~R. Brubaker}, {and} \bibinfo{person}{Danielle~Albers Szafir}.} \bibinfo{year}{2022}\natexlab{}.
\newblock \showarticletitle{Scholastic: {Graphical} {Human}-{AI} {Collaboration} for {Inductive} and {Interpretive} {Text} {Analysis}}. In \bibinfo{booktitle}{\emph{Proceedings of the 35th {Annual} {ACM} {Symposium} on {User} {Interface} {Software} and {Technology}}} \emph{(\bibinfo{series}{{UIST} '22})}. \bibinfo{publisher}{Association for Computing Machinery}, \bibinfo{address}{New York, NY, USA}, \bibinfo{pages}{1--12}.
\newblock
\showISBNx{978-1-4503-9320-1}
\href{https://doi.org/10.1145/3526113.3545681}{doi:\nolinkurl{10.1145/3526113.3545681}}


\bibitem[Horvitz(1999)]%
        {10.1145/302979.303030}
\bibfield{author}{\bibinfo{person}{Eric Horvitz}.} \bibinfo{year}{1999}\natexlab{}.
\newblock \showarticletitle{Principles of mixed-initiative user interfaces}. In \bibinfo{booktitle}{\emph{Proceedings of the SIGCHI Conference on Human Factors in Computing Systems}} (Pittsburgh, Pennsylvania, USA) \emph{(\bibinfo{series}{CHI '99})}. \bibinfo{publisher}{Association for Computing Machinery}, \bibinfo{address}{New York, NY, USA}, \bibinfo{pages}{159–166}.
\newblock
\showISBNx{0201485591}
\href{https://doi.org/10.1145/302979.303030}{doi:\nolinkurl{10.1145/302979.303030}}


\bibitem[Hou et~al\mbox{.}(2023)]%
        {10.1145/3544548.3581066}
\bibfield{author}{\bibinfo{person}{Yoyo Tsung-Yu Hou}, \bibinfo{person}{Wen-Ying Lee}, {and} \bibinfo{person}{Malte Jung}.} \bibinfo{year}{2023}\natexlab{}.
\newblock \showarticletitle{“Should I Follow the Human, or Follow the Robot?” — Robots in Power Can Have More Influence Than Humans on Decision-Making}. In \bibinfo{booktitle}{\emph{Proceedings of the 2023 CHI Conference on Human Factors in Computing Systems}} (Hamburg, Germany) \emph{(\bibinfo{series}{CHI '23})}. \bibinfo{publisher}{Association for Computing Machinery}, \bibinfo{address}{New York, NY, USA}, Article \bibinfo{articleno}{114}, \bibinfo{numpages}{13}~pages.
\newblock
\showISBNx{9781450394215}
\href{https://doi.org/10.1145/3544548.3581066}{doi:\nolinkurl{10.1145/3544548.3581066}}


\bibitem[Kang et~al\mbox{.}(2022a)]%
        {threddy}
\bibfield{author}{\bibinfo{person}{Hyeonsu Kang}, \bibinfo{person}{Joseph~Chee Chang}, \bibinfo{person}{Yongsung Kim}, {and} \bibinfo{person}{Aniket Kittur}.} \bibinfo{year}{2022}\natexlab{a}.
\newblock \showarticletitle{Threddy: {An} {Interactive} {System} for {Personalized} {Thread}-based {Exploration} and {Organization} of {Scientific} {Literature}}. In \bibinfo{booktitle}{\emph{Proceedings of the 35th {Annual} {ACM} {Symposium} on {User} {Interface} {Software} and {Technology}}} \emph{(\bibinfo{series}{{UIST} '22})}. \bibinfo{publisher}{Association for Computing Machinery}, \bibinfo{address}{New York, NY, USA}, \bibinfo{pages}{1--15}.
\newblock
\showISBNx{978-1-4503-9320-1}
\href{https://doi.org/10.1145/3526113.3545660}{doi:\nolinkurl{10.1145/3526113.3545660}}


\bibitem[Kang et~al\mbox{.}(2022b)]%
        {AnalogicalSearchEngine}
\bibfield{author}{\bibinfo{person}{Hyeonsu~B. Kang}, \bibinfo{person}{Xin Qian}, \bibinfo{person}{Tom Hope}, \bibinfo{person}{Dafna Shahaf}, \bibinfo{person}{Joel Chan}, {and} \bibinfo{person}{Aniket Kittur}.} \bibinfo{year}{2022}\natexlab{b}.
\newblock \showarticletitle{Augmenting {Scientific} {Creativity} with an {Analogical} {Search} {Engine}}.
\newblock \bibinfo{journal}{\emph{ACM Trans. Comput.-Hum. Interact.}} \bibinfo{volume}{29}, \bibinfo{number}{6} (\bibinfo{date}{Nov.} \bibinfo{year}{2022}), \bibinfo{pages}{57:1--57:36}.
\newblock
\showISSN{1073-0516}
\href{https://doi.org/10.1145/3530013}{doi:\nolinkurl{10.1145/3530013}}


\bibitem[Kang et~al\mbox{.}(2023)]%
        {synergi}
\bibfield{author}{\bibinfo{person}{Hyeonsu~B Kang}, \bibinfo{person}{Tongshuang Wu}, \bibinfo{person}{Joseph~Chee Chang}, {and} \bibinfo{person}{Aniket Kittur}.} \bibinfo{year}{2023}\natexlab{}.
\newblock \showarticletitle{Synergi: {A} {Mixed}-{Initiative} {System} for {Scholarly} {Synthesis} and {Sensemaking}}. In \bibinfo{booktitle}{\emph{Proceedings of the 36th {Annual} {ACM} {Symposium} on {User} {Interface} {Software} and {Technology}}} \emph{(\bibinfo{series}{{UIST} '23})}. \bibinfo{publisher}{Association for Computing Machinery}, \bibinfo{address}{New York, NY, USA}, \bibinfo{pages}{1--19}.
\newblock
\showISBNx{979-8-4007-0132-0}
\href{https://doi.org/10.1145/3586183.3606759}{doi:\nolinkurl{10.1145/3586183.3606759}}


\bibitem[Kumar et~al\mbox{.}(2024)]%
        {kumar2024human}
\bibfield{author}{\bibinfo{person}{Harsh Kumar}, \bibinfo{person}{Jonathan Vincentius}, \bibinfo{person}{Ewan Jordan}, {and} \bibinfo{person}{Ashton Anderson}.} \bibinfo{year}{2024}\natexlab{}.
\newblock \bibinfo{title}{Human Creativity in the Age of LLMs: Randomized Experiments on Divergent and Convergent Thinking}.
\newblock
\href{https://doi.org/10.48550/arXiv.2410.03703}{doi:\nolinkurl{10.48550/arXiv.2410.03703}}


\bibitem[Lee et~al\mbox{.}(2025)]%
        {lee2025impact}
\bibfield{author}{\bibinfo{person}{Hao-Ping~Hank Lee}, \bibinfo{person}{Advait Sarkar}, \bibinfo{person}{Lev Tankelevitch}, \bibinfo{person}{Ian Drosos}, \bibinfo{person}{Sean Rintel}, \bibinfo{person}{Richard Banks}, {and} \bibinfo{person}{Nicholas Wilson}.} \bibinfo{year}{2025}\natexlab{}.
\newblock \showarticletitle{The Impact of Generative AI on Critical Thinking: Self-Reported Reductions in Cognitive Effort and Confidence Effects From a Survey of Knowledge Workers}.
\newblock  (\bibinfo{year}{2025}).
\newblock


\bibitem[Lee and See(2004)]%
        {lee2004trust}
\bibfield{author}{\bibinfo{person}{John~D Lee} {and} \bibinfo{person}{Katrina~A See}.} \bibinfo{year}{2004}\natexlab{}.
\newblock \showarticletitle{Trust in automation: Designing for appropriate reliance}.
\newblock \bibinfo{journal}{\emph{Human factors}} \bibinfo{volume}{46}, \bibinfo{number}{1} (\bibinfo{year}{2004}), \bibinfo{pages}{50--80}.
\newblock


\bibitem[Lee et~al\mbox{.}(2024)]%
        {paperweaver}
\bibfield{author}{\bibinfo{person}{Yoonjoo Lee}, \bibinfo{person}{Hyeonsu~B Kang}, \bibinfo{person}{Matt Latzke}, \bibinfo{person}{Juho Kim}, \bibinfo{person}{Jonathan Bragg}, \bibinfo{person}{Joseph~Chee Chang}, {and} \bibinfo{person}{Pao Siangliulue}.} \bibinfo{year}{2024}\natexlab{}.
\newblock \showarticletitle{{PaperWeaver}: {Enriching} {Topical} {Paper} {Alerts} by {Contextualizing} {Recommended} {Papers} with {User}-collected {Papers}}. In \bibinfo{booktitle}{\emph{Proceedings of the 2024 {CHI} {Conference} on {Human} {Factors} in {Computing} {Systems}}} \emph{(\bibinfo{series}{{CHI} '24})}. \bibinfo{publisher}{Association for Computing Machinery}, \bibinfo{address}{New York, NY, USA}, \bibinfo{pages}{1--19}.
\newblock
\showISBNx{979-8-4007-0330-0}
\href{https://doi.org/10.1145/3613904.3642196}{doi:\nolinkurl{10.1145/3613904.3642196}}


\bibitem[Liu et~al\mbox{.}(2024)]%
        {CoQuest}
\bibfield{author}{\bibinfo{person}{Yiren Liu}, \bibinfo{person}{Si Chen}, \bibinfo{person}{Haocong Cheng}, \bibinfo{person}{Mengxia Yu}, \bibinfo{person}{Xiao Ran}, \bibinfo{person}{Andrew Mo}, \bibinfo{person}{Yiliu Tang}, {and} \bibinfo{person}{Yun Huang}.} \bibinfo{year}{2024}\natexlab{}.
\newblock \showarticletitle{How AI Processing Delays Foster Creativity: Exploring Research Question Co-Creation with an LLM-based Agent}. In \bibinfo{booktitle}{\emph{Proceedings of the 2024 CHI Conference on Human Factors in Computing Systems}} (Honolulu, HI, USA) \emph{(\bibinfo{series}{CHI '24})}. \bibinfo{publisher}{Association for Computing Machinery}, \bibinfo{address}{New York, NY, USA}, Article \bibinfo{articleno}{17}, \bibinfo{numpages}{25}~pages.
\newblock
\showISBNx{9798400703300}
\href{https://doi.org/10.1145/3613904.3642698}{doi:\nolinkurl{10.1145/3613904.3642698}}


\bibitem[Luo et~al\mbox{.}(2025)]%
        {luo2025llm4srsurveylargelanguage}
\bibfield{author}{\bibinfo{person}{Ziming Luo}, \bibinfo{person}{Zonglin Yang}, \bibinfo{person}{Zexin Xu}, \bibinfo{person}{Wei Yang}, {and} \bibinfo{person}{Xinya Du}.} \bibinfo{year}{2025}\natexlab{}.
\newblock \bibinfo{title}{LLM4SR: A Survey on Large Language Models for Scientific Research}.
\newblock
\showeprint[arxiv]{2501.04306}~[cs.CL]
\urldef\tempurl%
\url{https://arxiv.org/abs/2501.04306}
\showURL{%
\tempurl}


\bibitem[{OpenAI}(2022)]%
        {openai2022chatgpt}
\bibfield{author}{\bibinfo{person}{{OpenAI}}.} \bibinfo{year}{2022}\natexlab{}.
\newblock \bibinfo{title}{Introducing ChatGPT}.
\newblock
\urldef\tempurl%
\url{https://openai.com/index/chatgpt/}
\showURL{%
\tempurl}
\newblock
\shownote{Accessed: 2025-02-20}.


\bibitem[Overney et~al\mbox{.}(2024)]%
        {sensemate}
\bibfield{author}{\bibinfo{person}{Cassandra Overney}, \bibinfo{person}{Belén Saldías}, \bibinfo{person}{Dimitra Dimitrakopoulou}, {and} \bibinfo{person}{Deb Roy}.} \bibinfo{year}{2024}\natexlab{}.
\newblock \showarticletitle{{SenseMate}: {An} {Accessible} and {Beginner}-{Friendly} {Human}-{AI} {Platform} for {Qualitative} {Data} {Analysis}}. In \bibinfo{booktitle}{\emph{Proceedings of the 29th {International} {Conference} on {Intelligent} {User} {Interfaces}}} \emph{(\bibinfo{series}{{IUI} '24})}. \bibinfo{publisher}{Association for Computing Machinery}, \bibinfo{address}{New York, NY, USA}, \bibinfo{pages}{922--939}.
\newblock
\showISBNx{979-8-4007-0508-3}
\href{https://doi.org/10.1145/3640543.3645194}{doi:\nolinkurl{10.1145/3640543.3645194}}


\bibitem[Palani et~al\mbox{.}(2023)]%
        {Relatedly}
\bibfield{author}{\bibinfo{person}{Srishti Palani}, \bibinfo{person}{Aakanksha Naik}, \bibinfo{person}{Doug Downey}, \bibinfo{person}{Amy~X. Zhang}, \bibinfo{person}{Jonathan Bragg}, {and} \bibinfo{person}{Joseph~Chee Chang}.} \bibinfo{year}{2023}\natexlab{}.
\newblock \showarticletitle{Relatedly: Scaffolding Literature Reviews with Existing Related Work Sections}. In \bibinfo{booktitle}{\emph{Proceedings of the 2023 CHI Conference on Human Factors in Computing Systems}} (Hamburg, Germany) \emph{(\bibinfo{series}{CHI '23})}. \bibinfo{publisher}{Association for Computing Machinery}, \bibinfo{address}{New York, NY, USA}, Article \bibinfo{articleno}{742}, \bibinfo{numpages}{20}~pages.
\newblock
\showISBNx{9781450394215}
\href{https://doi.org/10.1145/3544548.3580841}{doi:\nolinkurl{10.1145/3544548.3580841}}


\bibitem[Pang et~al\mbox{.}(2025)]%
        {pang2025understandingllmificationchiunpacking}
\bibfield{author}{\bibinfo{person}{Rock~Yuren Pang}, \bibinfo{person}{Hope Schroeder}, \bibinfo{person}{Kynnedy~Simone Smith}, \bibinfo{person}{Solon Barocas}, \bibinfo{person}{Ziang Xiao}, \bibinfo{person}{Emily Tseng}, {and} \bibinfo{person}{Danielle Bragg}.} \bibinfo{year}{2025}\natexlab{}.
\newblock \bibinfo{title}{Understanding the LLM-ification of CHI: Unpacking the Impact of LLMs at CHI through a Systematic Literature Review}.
\newblock
\showeprint[arxiv]{2501.12557}~[cs.HC]
\urldef\tempurl%
\url{https://arxiv.org/abs/2501.12557}
\showURL{%
\tempurl}


\bibitem[Parasuraman et~al\mbox{.}(2000)]%
        {parasuraman2000model}
\bibfield{author}{\bibinfo{person}{Raja Parasuraman}, \bibinfo{person}{Thomas~B Sheridan}, {and} \bibinfo{person}{Christopher~D Wickens}.} \bibinfo{year}{2000}\natexlab{}.
\newblock \showarticletitle{A model for types and levels of human interaction with automation}.
\newblock \bibinfo{journal}{\emph{IEEE Transactions on systems, man, and cybernetics-Part A: Systems and Humans}} \bibinfo{volume}{30}, \bibinfo{number}{3} (\bibinfo{year}{2000}), \bibinfo{pages}{286--297}.
\newblock


\bibitem[Passi and Vorvoreanu(2022)]%
        {passi2022overreliance}
\bibfield{author}{\bibinfo{person}{Samir Passi} {and} \bibinfo{person}{Mihaela Vorvoreanu}.} \bibinfo{year}{2022}\natexlab{}.
\newblock \showarticletitle{Overreliance on AI literature review}.
\newblock \bibinfo{journal}{\emph{Microsoft Research}} (\bibinfo{year}{2022}).
\newblock


\bibitem[Pirolli and Card(2005)]%
        {pirolli2005sensemaking}
\bibfield{author}{\bibinfo{person}{Peter Pirolli} {and} \bibinfo{person}{Stuart Card}.} \bibinfo{year}{2005}\natexlab{}.
\newblock \showarticletitle{The sensemaking process and leverage points for analyst technology as identified through cognitive task analysis}. In \bibinfo{booktitle}{\emph{Proceedings of international conference on intelligence analysis}}, Vol.~\bibinfo{volume}{5}. McLean, VA, USA, \bibinfo{pages}{2--4}.
\newblock


\bibitem[Pu et~al\mbox{.}(2024)]%
        {IdeaSynth}
\bibfield{author}{\bibinfo{person}{Kevin Pu}, \bibinfo{person}{K.~J.~Kevin Feng}, \bibinfo{person}{Tovi Grossman}, \bibinfo{person}{Tom Hope}, \bibinfo{person}{Bhavana~Dalvi Mishra}, \bibinfo{person}{Matt Latzke}, \bibinfo{person}{Jonathan Bragg}, \bibinfo{person}{Joseph~Chee Chang}, {and} \bibinfo{person}{Pao Siangliulue}.} \bibinfo{year}{2024}\natexlab{}.
\newblock \bibinfo{title}{IdeaSynth: Iterative Research Idea Development Through Evolving and Composing Idea Facets with Literature-Grounded Feedback}.
\newblock
\showeprint[arxiv]{2410.04025}~[cs.HC]
\urldef\tempurl%
\url{https://arxiv.org/abs/2410.04025}
\showURL{%
\tempurl}


\bibitem[Russell et~al\mbox{.}(1993)]%
        {russell1993cost}
\bibfield{author}{\bibinfo{person}{Daniel~M Russell}, \bibinfo{person}{Mark~J Stefik}, \bibinfo{person}{Peter Pirolli}, {and} \bibinfo{person}{Stuart~K Card}.} \bibinfo{year}{1993}\natexlab{}.
\newblock \showarticletitle{The cost structure of sensemaking}. In \bibinfo{booktitle}{\emph{Proceedings of the INTERACT'93 and CHI'93 conference on Human factors in computing systems}}. \bibinfo{pages}{269--276}.
\newblock


\bibitem[Shneiderman(2007)]%
        {shneiderman2007creativity}
\bibfield{author}{\bibinfo{person}{Ben Shneiderman}.} \bibinfo{year}{2007}\natexlab{}.
\newblock \showarticletitle{Creativity support tools: accelerating discovery and innovation}.
\newblock \bibinfo{journal}{\emph{Commun. ACM}} \bibinfo{volume}{50}, \bibinfo{number}{12} (\bibinfo{year}{2007}), \bibinfo{pages}{20--32}.
\newblock


\bibitem[Shneiderman(2020)]%
        {shneiderman2020human}
\bibfield{author}{\bibinfo{person}{Ben Shneiderman}.} \bibinfo{year}{2020}\natexlab{}.
\newblock \showarticletitle{Human-centered artificial intelligence: Reliable, safe \& trustworthy}.
\newblock \bibinfo{journal}{\emph{International Journal of Human--Computer Interaction}} \bibinfo{volume}{36}, \bibinfo{number}{6} (\bibinfo{year}{2020}), \bibinfo{pages}{495--504}.
\newblock


\bibitem[Suh et~al\mbox{.}(2023)]%
        {Sensecape}
\bibfield{author}{\bibinfo{person}{Sangho Suh}, \bibinfo{person}{Bryan Min}, \bibinfo{person}{Srishti Palani}, {and} \bibinfo{person}{Haijun Xia}.} \bibinfo{year}{2023}\natexlab{}.
\newblock \showarticletitle{Sensecape: Enabling Multilevel Exploration and Sensemaking with Large Language Models}. In \bibinfo{booktitle}{\emph{Proceedings of the 36th Annual ACM Symposium on User Interface Software and Technology}} (San Francisco, CA, USA) \emph{(\bibinfo{series}{UIST '23})}. \bibinfo{publisher}{Association for Computing Machinery}, \bibinfo{address}{New York, NY, USA}, Article \bibinfo{articleno}{1}, \bibinfo{numpages}{18}~pages.
\newblock
\showISBNx{9798400701320}
\href{https://doi.org/10.1145/3586183.3606756}{doi:\nolinkurl{10.1145/3586183.3606756}}


\bibitem[Zheng et~al\mbox{.}(2024)]%
        {disciplink}
\bibfield{author}{\bibinfo{person}{Chengbo Zheng}, \bibinfo{person}{Yuanhao Zhang}, \bibinfo{person}{Zeyu Huang}, \bibinfo{person}{Chuhan Shi}, \bibinfo{person}{Minrui Xu}, {and} \bibinfo{person}{Xiaojuan Ma}.} \bibinfo{year}{2024}\natexlab{}.
\newblock \showarticletitle{DiscipLink: Unfolding Interdisciplinary Information Seeking Process via Human-AI Co-Exploration}. In \bibinfo{booktitle}{\emph{Proceedings of the 37th {Annual} {ACM} {Symposium} on {User} {Interface} {Software} and {Technology}}} \emph{(\bibinfo{series}{{UIST} '24})}. \bibinfo{publisher}{Association for Computing Machinery}, \bibinfo{address}{New York, NY, USA}, \bibinfo{pages}{1--20}.
\newblock
\showISBNx{979-8-4007-0628-8}
\href{https://doi.org/10.1145/3654777.3676366}{doi:\nolinkurl{10.1145/3654777.3676366}}


\end{thebibliography}
